\documentclass[%
reprint,
showpacs,preprintnumbers,
 amsmath,amssymb,
 aps,
prl,
]{revtex4-1}

\usepackage{graphicx}
\usepackage{dcolumn}
\usepackage{bm}

\begin{document}


\title{Localization of low-frequency oscillations in single-walled carbon nanotubes}

\author{V.V. Smirnov}%
\email{vvs@polymer.chph.ras.ru}
\author{D.S.Shepelev}
\author{L.I. Manevitch}
 
\affiliation{%
 Institute of Chemical Physics, RAS, Moscow, Russia\\
119991, 4 Kosygin str., Moscow, Russia
}%

\date{\today}

\begin{abstract}
New phenomenon of a weak energy localization of optical low-frequency oscillations in carbon nanotubes (CNT) is analytically predicted in the framework of continuum shell theory. This phenomenon takes place for CNTs of finite length with medium aspect ratio. The origin of localization is clarified by means of the concept of Limiting Phase Trajectory, and the analytical results are confirmed by the MD simulation of simply supported CNT.
\end{abstract}

\maketitle




 From a modern point of view, carbon nanotubes are twice the exciting subject of scientific researches. On the one hand, they are associated with great hopes for creation of super-small and ultra-fast electronic and electromechanical devices \cite{CLi03, Sazonova2004, Peng2006,  Anantram2006}. On the other hand, they are quasi-one-dimensional objects that allow to check out some of the funamentals of modern solid-state physics. In particular, variuos computational and in-situ measurements of thermoconductivity of CNT \cite{Berber00, Wang06, Mingo05,BLi05, Savin09} are directly related  with the problem of finiteness of thermoconductivity of one-dimensional anharmonic lattices. This problem has been formulated more than fifty years ago in the famous work by Fermi, Pasta and Ulam \cite{FPU}. The wide-area study of nonlinear lattices dynamics led to discovery of new class of elementary excitations - solitons, the main feature of those is the self-localization in the homogeneous lattices \cite{Scott, Peyrard}. From the point of view of energy trasfer, solitons take place dual role. Being very effective energy carriers they also provide the effective scattering of small-amplitude phonons \cite{Gendelman2000, Lepri2003, Lepri2003_2, Zhang2013, NLi2010, Zhong2012, BLi2005}. From the mathematical point of view, the solution like a breather exists only in the infinite systems with a continuous spectrum, while the nanoscale objects can be rather considered as finite ones. In such a case the problem of definition of nonlinear localized excitations has significant pecularities.
 
It was recently shown \cite{VVS2010, DAN2010, VVS2011} that the finite systems of weakly coupled oscillators exhibit strongly modulated non-stationary oscillations characterized by the maximum possible energy exchange between the groups of the oscillators or the maximum energy transfer from the external source of energy to the system \cite{Man07}. The solutions describing these processes are referred to as Limiting Phase Trajectories (LPTs). The development and the use of the analytical framework based on the LPT concept is motivated by the fact that resonant non-stationary processes occurring in a broad variety of finite dimensional physical models are beyond the well-known paradigm of nonlinear normal modes (NNMs), fully justified only for quasi-stationary processes and non-stationary processes in non-resonant case. While the NNMs approach has been proved to be an effective tool for the analysis of instability and bifurcations of stationary processes (see, e.g., \cite{Vak96}), the use of the LPTs concept provides the adequate procedures for studying strongly modulated regimes as well as the transitions to energy localization and chaotic behavior \cite{VVS2010}. Such an approach clarifies also the physical nature of the breathers formation in infinite discrete or continuum systems. This letter covers the description of energy exchange and its localization in nanotubes, based on this approach in combination with the thin shell model of CNT.

Calculation of the phonon spectra of CNT is a subject of a  number of studies, many of which are based on the continuum model of a single-walled carbon nanotubes as a thin elastic shell \cite{Dresselhaus00, CYWang04, Duan07, Liew07, Gupta10, Elishakoff09, Arghavan11, Silvestre11, Silvestre12, Ghavanloo12, Chang10, Shi08, Shi09, Soltani11}. Beginning with the pioneering work of Jacobson \cite{PRL96}, this approach is actively exploited in the description of the various mechanical properties of CNTs and its adequacy was repeatedly checked by comparison to  experimental data as well as to data of molecular dynamic simulation \cite {CYWang04, Silvestre11, Silvestre12, Ghavanloo12}. However, the theory of thin shells allows an analytical solution only in exceptional, stationary cases \cite{Amabili08, Kumar13} even in the framework of linear approximation. To the best of our knowledge, nonlinear dynamic processes in CNTs were analytically studied only on the base of a simplest modal analysis (axially symmetric nonlinear normal mode  and its parametric instability)  \cite {Shi08, Shi09, Soltani11}.  

\begin{figure}

\centering{
 \includegraphics[width=60 mm]{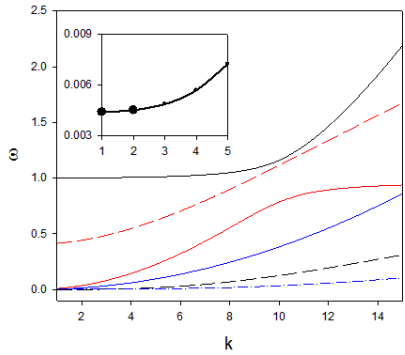}
}
\caption{
The CNT spectrum in according to the exact Sanders-Koiter thin shell theory: solid curves correspond to circumferential number $n=0$, dashed ones - to $n=1$ and dot-dashed curve - to $n=2$. The insert shows the small wave number part of CFM branch and two interacting modes are marked by black cicles. All frequencies $\omega$ are measured in the dimensionless units and the longitudinal wave number $k$ - in the numbers of half-wave along the CNT axes.}
\label{fig:CNT2chain}

\end{figure}


There are two optical-type vibration branches in the CNT spectrum (fig. \ref{fig:CNT2chain}): the first one is the well-known Radial Breathing Mode (RBM), which is associated with azimuthal wave number $n=0$ and corresponds to uniform radial expansion-compression. The second mode is specified by $n=2$, and the main deformation is a deviation of CNT cross-section from initial circular one (Circumferential Flexure Mode - CFM) \cite{Rao97}. One should note that CFM is the lowest optical mode in the CNT spectrum. In this Letter we discuss the model of low-frequency optical vibrations, based on the new version of theory of thin shells (see Supplementary materials), starting from the Sanders-Koiter approximation \cite{Amabili08}.

%
%
 %


Let us consider the oscillations relating to CFM branch.  Chosen branch of the spectrum (CFM) is characterized by relative smallness of the ring and shear  deformations - $\varepsilon_{\theta}$ and $\varepsilon_{\xi \theta} $ ($\xi$ is the dimensionless coordinate along the CNT axis and $\theta$ is the azimuthal angle), which are supposed to be "the small differences of large values". Consequently, the tangential and longitudinal components of the displacement can be expressed via radial component and the energies of ring and shear deformations can be neglected. Thus, one gets the equation of motion for a single variable characterizing the radial displacement (see Supplementary materials).


%


The resulting non-linear equation provides the analysis of both linear vibration spectrum  and nonlinearity effect on frequency depending on the boundary conditions. The significant feature of the spectrum obtained, as well as of all optical type spectra, is the eigenvalues crowding near the spectrum edges, regardless of the type of boundary conditions. The spectrum density grows with increasing the length of the CNT. 
This means that even in the linear approximation one should take into account the resonance of normal modes, which manifests primarily at the edges of the spectrum. This effect is of a general nature, therefore we show the main results on the example of the simple boundary conditions - simply supported CNTs. 
In this case  the radial and circumferential displacements as well as longitudinal bending moment and axial force are equal to zero at the CNT edges.  According to the hypothesis suggested above, we restrict the analysis to the two-mode approximation, assuming that the rest of the spectrum is separated by a gap wide enough and does not make a appreciable contribution to the resonant interaction between the lowest (by frequency) modes (one should note that the reqiured smallness of frequency gap is achieved for the CNTs with aspect ratio - the ratio of CNT length to its radus - $\lambda\geq 15$). Thus, we represent the solution of the nonlinear equation  as

\begin{equation}
w(\xi,\theta,t)=(f_{1}(t) sin(\pi \xi)+f_{2}(t) sin(2\pi \xi))cos(2\theta),
\label{mode2}
\end{equation}

where $w$ is radial displacement of the shell.

Assuming that the main effect of non-linearity manifests primarily in the time dependence of the amplitudes of $f_{1}$ and $f_{2}$, one can use the Galerkin method to obtain the equations for them. However, the problem remains too complicated for analytical study. We used the asymptotic analysis to obtain the equations describing the dynamics of low-frequency normal modes. 
 


Applying the asymptotic expansion in terms of the natural small parameter, which is the relative difference between the eigenvalues of low-frequency modes, and the method of multiple scales \cite{VVS2010}, we can obtain the equations for the complex amplitudes in the main approximation:

\begin{equation}
\label{main_app}
\begin{split}
i \frac{\partial \chi_{1}}{\partial \tau}+a_{1}|\chi_{1}|^{2}\chi_{1}+a_{3}|\chi_{2}|^{2}\chi_{1}+a_{4}\chi_{2}^{2}\chi_{1}^{*}=0 \\
i \frac{\partial \chi_{2}}{\partial \tau}+\omega_{1}\chi_{2}+a_{3}|\chi_{2}|^{2}\chi_{1}+a_{2}|\chi_{2}|^{2}\chi_{2}+a_{4}\chi_{1}^{2}\chi_{2}^{*}=0
\end{split}
\end{equation}

where amplitudes $\chi_{j}$  relate with $f_{j}$ by

 \begin{equation}
\label{asympt_1}
\begin{split}
\varepsilon\chi_{j}=\frac{1}{\sqrt{2 \omega_{j}}}(\frac{\partial f_{j}}{\partial \tau}+i \omega_{j} f_{j}) e^{-\omega_{j} t} \\
\varepsilon=\frac{\sqrt{\omega_{2}^{2}-\omega_{1}^{2}}}{\omega_{1}}
\end{split}
\end{equation}

The coefficients $a_{j}$ are determined by the geometry of CNT and the Poisson ratio, and $\omega_{j}$ are natural frequencies of CNT vibration.
The variable $\tau$ in equations (\ref{main_app}) is a slow time ($\tau=\varepsilon^{2} t$). It can be shown that the complex amplitudes $\chi_{j}$ depend only on this time. The procedure for deriving the equations (\ref{main_app}) has been described in detail in \cite{VVS2010} for discrete $\beta-$FPU lattice. In the considered case, equations (\ref{main_app}) correspond to Hamiltonian

\begin{equation}
\label{Hamilton1}
\begin{split}
 H=\omega_{1} |\chi_{2}|^{2}+a_{1}|\chi_{1}|^4+a_{2}|\chi_{2}|^{4}+a_{3}|\chi_{1}|^{2}|\chi_{2}|^{2}+  \\ 
 a_{4}(\chi_{1}^{2}\chi_{2}^{*2}+\chi_{1}^{*2}\chi_{2}^{2}) .
\end{split}
\end{equation}

Equations (\ref{main_app}), besides the obvious energy integral (\ref{Hamilton1}), possess another integral (occupation number integral in quantum-mechanical terminology):

\begin{equation}
\label{Occupation}
X=|\chi_{1}|^{2}+|\chi_{2}|^2
\end{equation}

 However, as it was shown in \cite{VVS2010}, the modal analysis becomes non-adequate at the resonance conditions. Therefore we introduce new variables as a linear combination of normal modes, which preserves the integral $X$:

\begin{equation}
\label{psi}
\psi_{1}=\frac{1}{\sqrt{2}}(\chi_{1}+\chi_{2}); \psi_{2}=\frac{1}{\sqrt{2}}(\chi_{1}-\chi_{2}).
\end{equation}
 
The new variables describe the dynamics of some parts of the CNT or some groups of the particles in the effective discrete one-dimensional chain (see Supplementary materials)  \cite{VVS2010, DAN2010, VVS2011}. 
Really, considering the distribution of energy along the nanotube one can see that such a linear combination of normal modes describes a predominant energy concentration in certain region of the CNT, while the other part of CNT has a lower density of the energy. Because of small difference between frequencies of the modes, the selected  parts of CNT demonstrate some coherent behavior similar to beating in the system of two weakly coupled oscillators. 
Therefore we can consider these regions as new large-scale elementary blocks, which can be described as unique elements of the system. Such blocks will be identified as "effective particles", which were introduced for discrete nonlinear lattices in  \cite{VVS2010}.  The existence of integral of motion (\ref{Occupation}) allows to reduce the phase space of the system up to 2 dimension. In this reduced phase space the first variable ($\theta$) specifies the relative amplitudes of effective particles and the second one ($\delta$) corresponds to the phase shift between them.

\begin{equation}
\label{Angle1}
\psi_{1}=\sqrt{X}\cos(\theta)e^{-i\delta/2} ;  
\psi_{2}=\sqrt{X}\sin(\theta)e^{i\delta/2}
\end{equation}

Now one can rewrite the Hamilton function (\ref{Hamilton1}) in the terms of "angle" variables $\theta$ and $\delta$:

\begin{equation}
\begin{split}
H =\frac{X}{16} \{ 8 a_{0} (1- \cos (\delta ) \sin (2 \theta ))+ \\  X [2 a_{4} ( 2 \cos ^{2}(\delta )-(\cos (2 \delta )-3) \cos (4 \theta ) )+ \\  a_{3} (2 \cos ^{2}(\delta ) \cos (4 \theta )  -\cos (2 \delta )+3 )+ \\  4 a_{2} (\cos (\delta ) \sin (2 \theta )-1)^{2}+ 
 4 a_{1} (\cos (\delta ) \sin (2 \theta )+1)^{2} ] \}
\end{split}
\label{Hamilton2}
\end{equation}

\begin{figure}
\noindent
\centering{
\includegraphics[width=40mm]{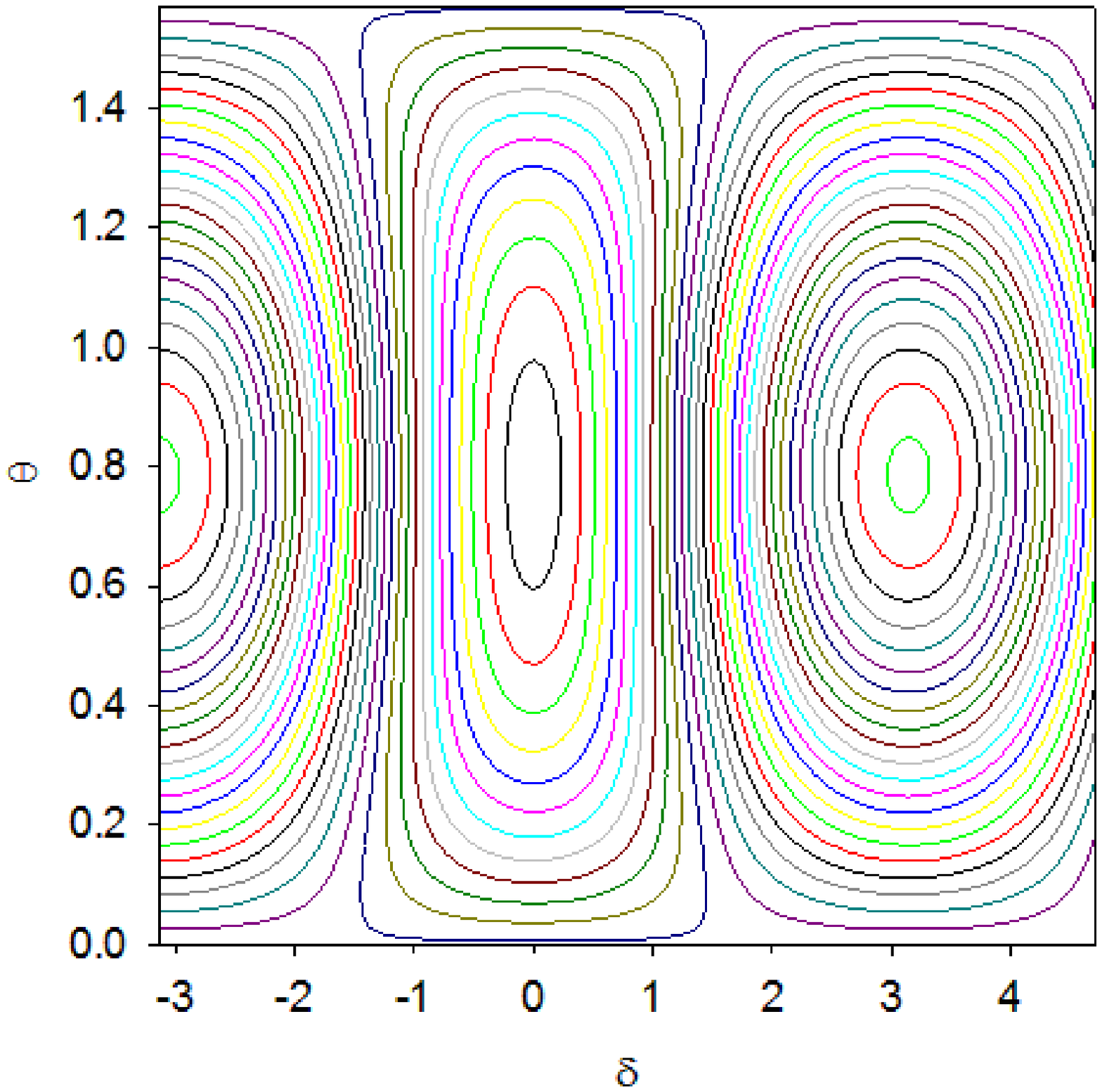} 
\includegraphics[width=40mm]{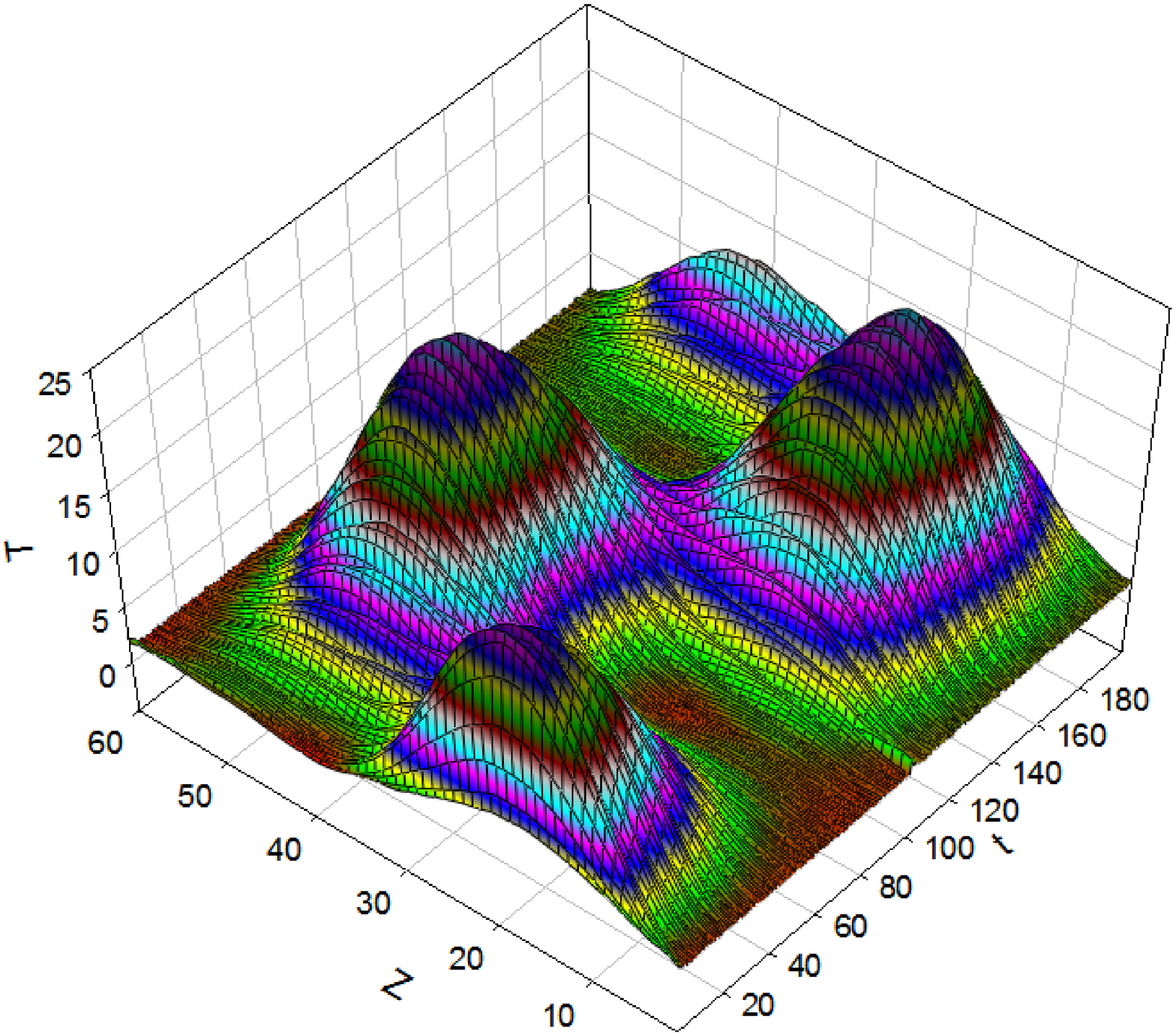}
}

\caption{Left panel: phase portrait of the system with low level of excitation. Right panel: the energy distribution along the CNT during the MD-simulation with the excitation, that corresponded to phase portrait in the left panel.}
\label{fig:smallX}

\end{figure}

A typical phase portrait of the system with Hamiltonian (\ref{Hamilton2}) for small values of $X$ is shown in Figure \ref{fig:smallX}(a). Two steady states with $\theta=\pi/4$ and $\delta=0$ and $\delta=\pi$ correspond to normal modes $\chi_{1}$ and $\chi_{2}$. The trajectories surrounding these stationary points describe the dynamics of 
effective particles, i.e. reflect an evolution of variables $\psi_{1}$ and $\psi_{2}$. Values $\theta=0$ and $\theta=\pi/2$ correspond
to energy concentration on the  "effective particles". In these states, the energy distribution along the CNT is the most non-uniform  one along the axis of the nanotube. 
However, it can be seen that these states belong to phase trajectory maximally distant from the stationary points. Consequently,  the motion along the trajectory is accompanied with the energy transfer from one part of CNT to another one. Such a trajectory was denoted as LPT - Limiting Phase Trajectory because it encircles the domain of attraction of the stationary points at maximal distance. The motion along LPT  is similar to beats in a system of two oscillators, and the parts of CNT  play the role of "effective particles". 
This process is well seen in figure \ref{fig:smallX}(b), where the energy distribution along CNT during the MD-simulation of the zigzag CNT oscillations is shown. The parameters of CNT: the length $L= 25.4$ nm, radius $R=0.79$ nm, the boundary conditions correspond to the simply supported nanotube. The initial temperature of nanotube was equal to 1.0 K and initial velocity field corresponded to the linear combination of two low-frequency modes (\ref{mode2}). To show the energy in figure \ref{fig:smallX}(b), the considered CNT  was  separated into 60 "elementary rings", each of them contained 40 carbon atoms.

\begin{figure}
\centering{

a \includegraphics[width=40mm]{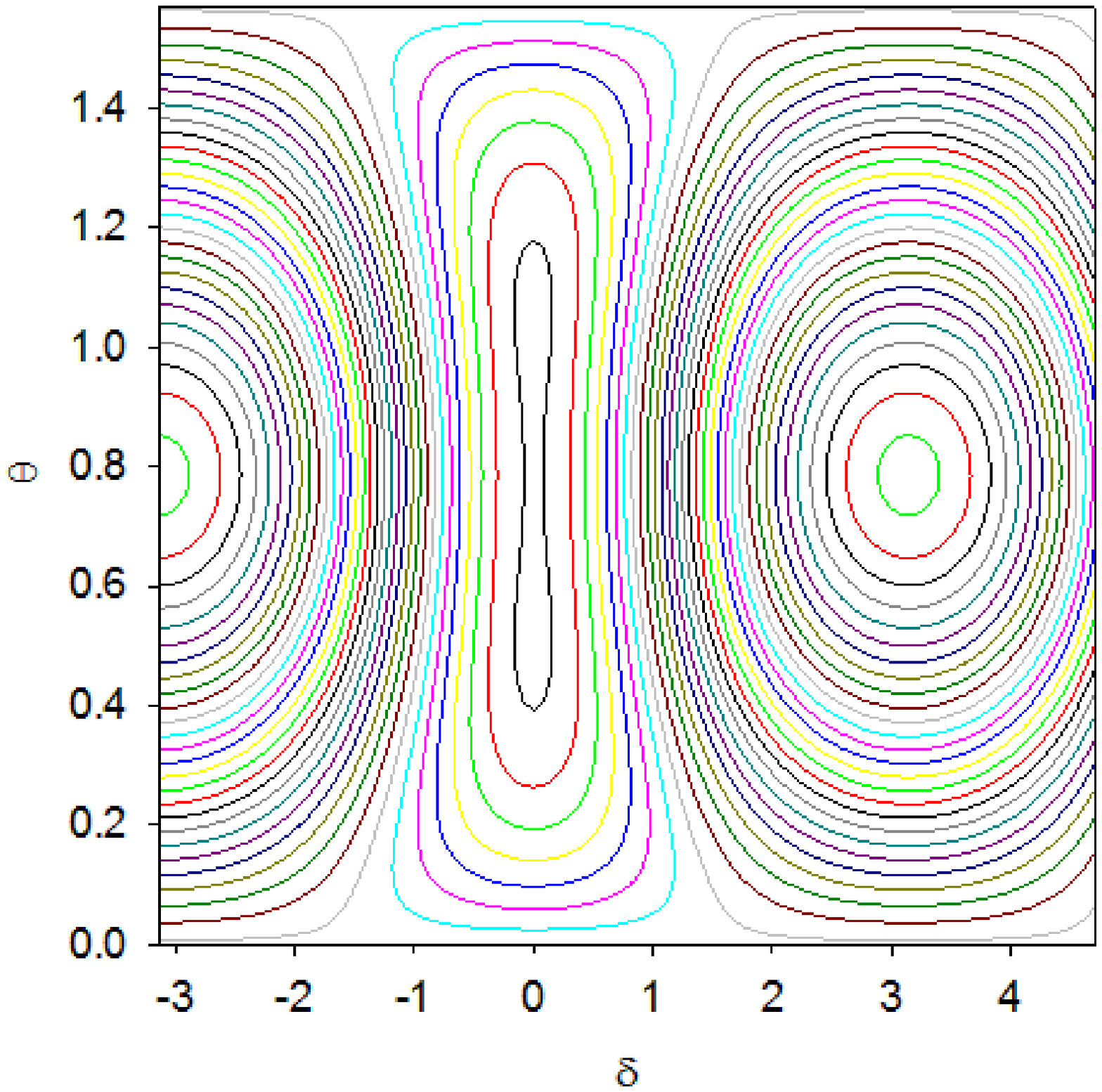}  b \includegraphics[width=40mm]{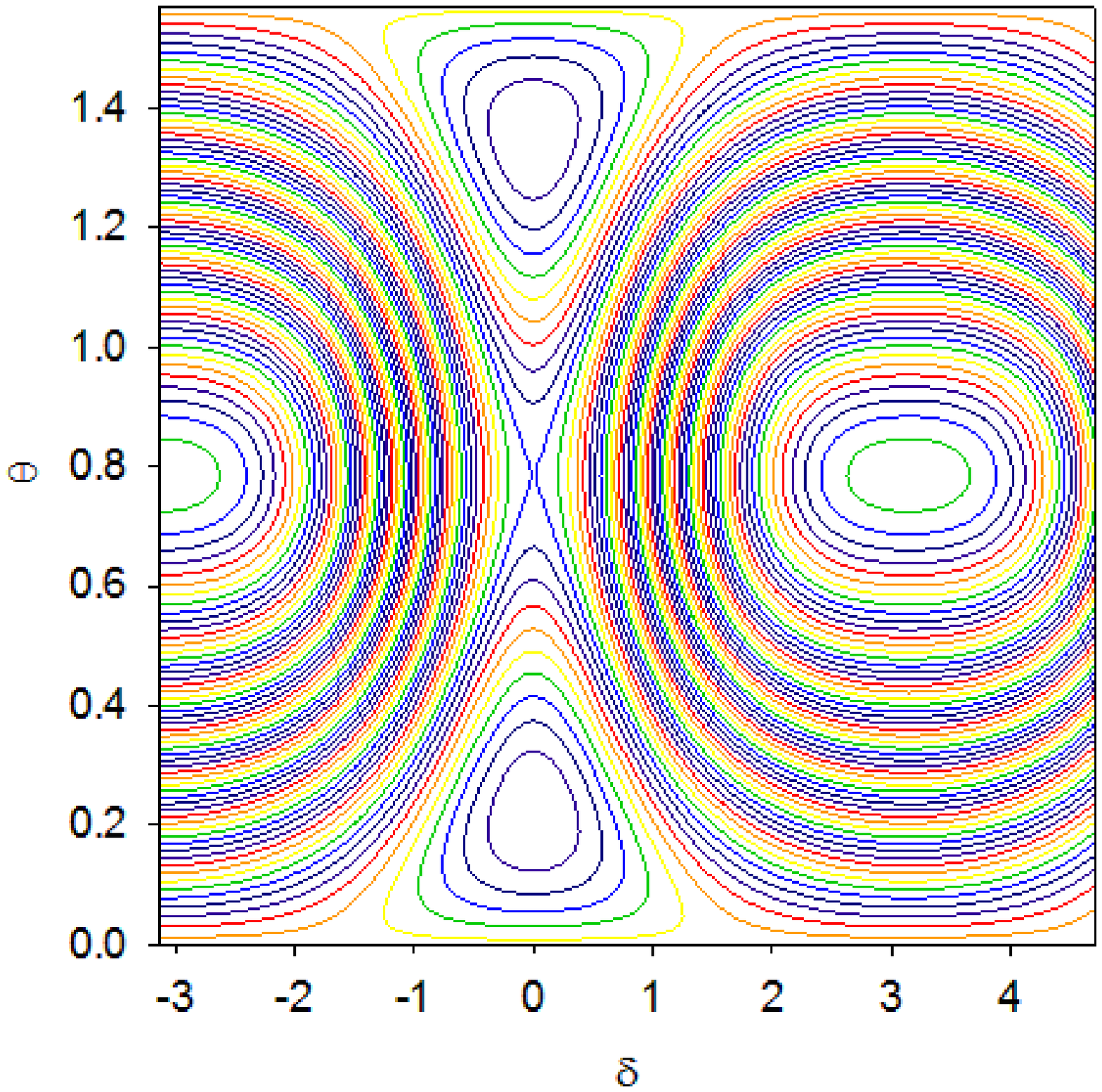} \\
c \includegraphics[width=40mm]{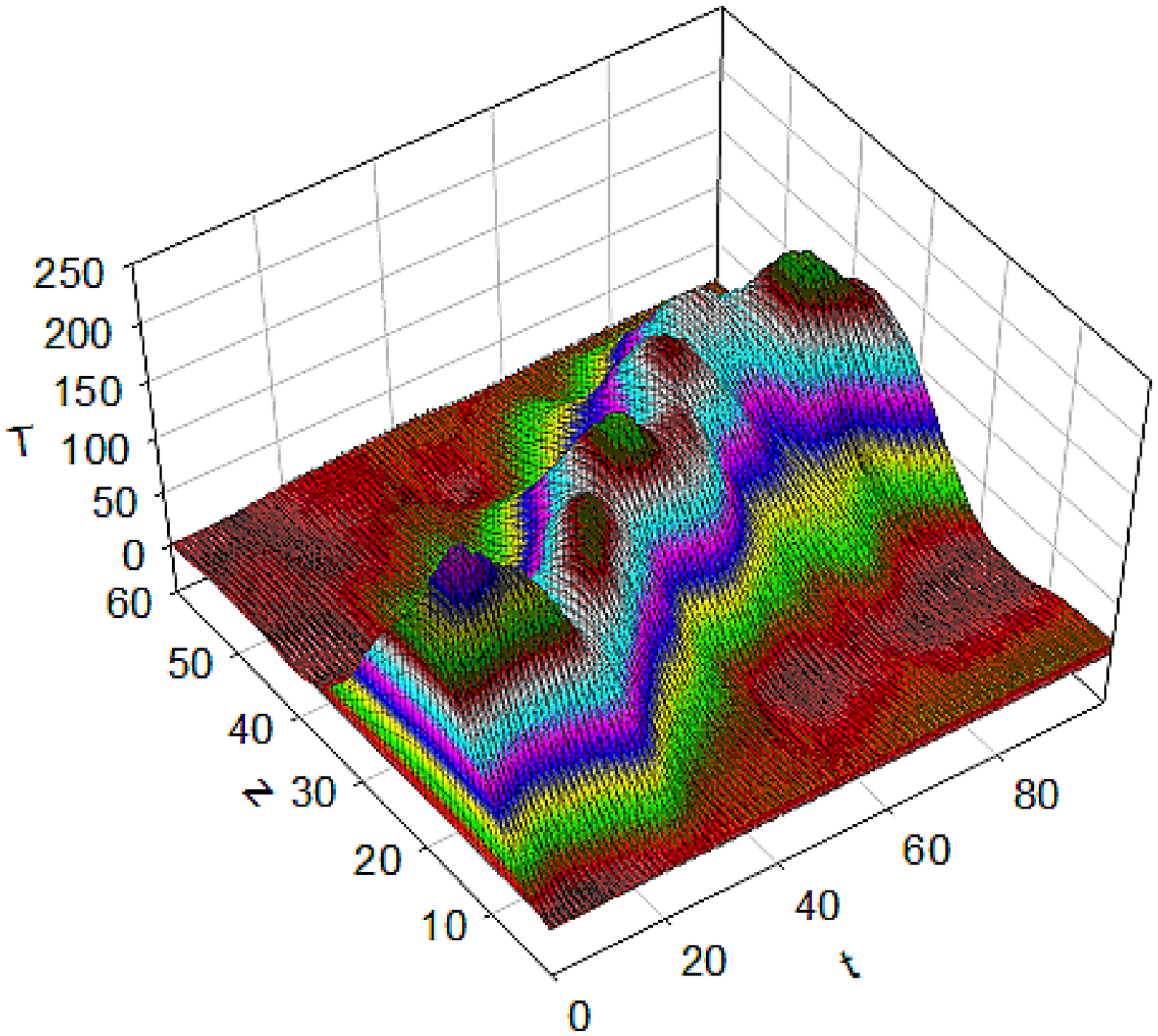}  d \includegraphics[width=40mm]{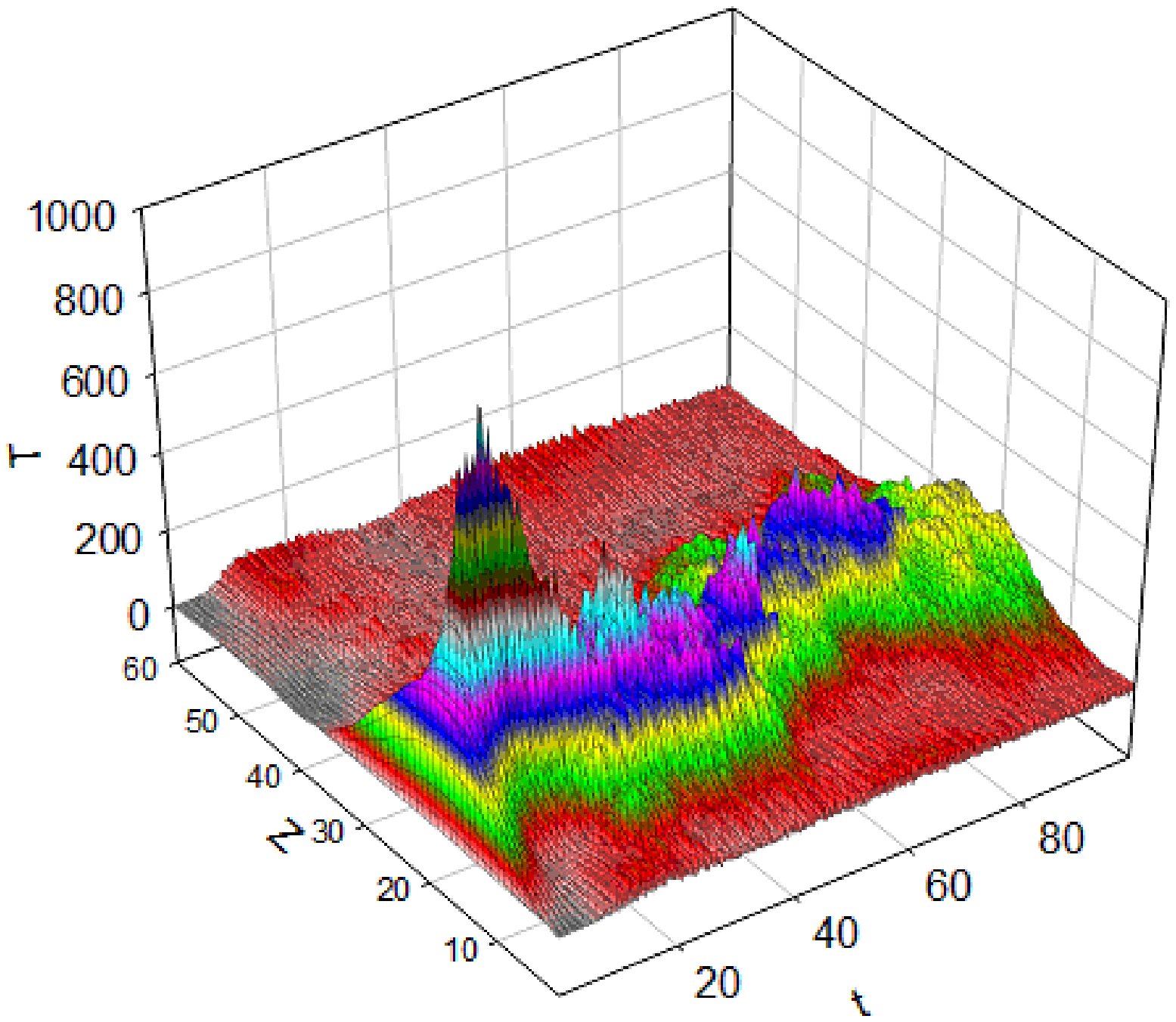} 
}
\caption{The phase portraits (a, b) and energy surfaces (c, d) at large excitation levels for CNTs with aspect ratio $\lambda=32$; left panel - $X=0.15$; right panel - $ X=0.25$. Energy is measured in Kelvin and time - in picoseconds. The results of MD-simulations are shown in the panels (c-d).}
\label{fig:largeX}


\end{figure}

What does occur when the parameter X grows? Figures \ref{fig:largeX}(a-b) show two phase portraits corresponding to large values of $X$. It can be seen that the topology of these portraits essentially differs from that in fig. \ref{fig:smallX}(a). Namely, except the initial stationary points corresponding to the original normal modes, the additional steady states arise, that is the result of instability of the lowest frequency normal mode.

The new steady states, as long as they are  close to the "parent"  normal mode, describe only a NNMs with relatively small energy excess  in one part of the CNT. 
But it is most importantly that the possibility of complete energy exchange between the CNT parts is preserved. With the growth of the parameter X new stationary states depart further from the original normal mode and, consequently, the separatrix loop is expanded. At some point, the separatrix becomes so expanded that it merges with the LPT. This threshold corresponds to the complete energy localization since any path, starting at the upper half of the phase plane, cannot reach the bottom one, and vice versa. Fig. \ref{fig:largeX} (a, b) illustrates this transformation of the phase plane. The instability threshold for the CNT with aspect ratio 32 is $X_{i}=0.08$ and the localization threshold is $X_{loc}=0.17$. The energy distribution along $z$-axis of CNT obtained by the MD simulation is shown in the panels (c-d) of figure \ref{fig:largeX}. One can see that near the localization threshold the beating which is shown in figure \ref{fig:smallX}(b) transforms into long-time flow of the energy from one part of CNT to another one (fig. \ref{fig:largeX}(c)). Fig. \ref{fig:largeX}(d) shows the energy distribution at excitation level, which exceeds the localization threshold significantly. One can see that no energy flow along the CNT occurs. In spite of the fact that the energy profile during the simulation differs significantly from initial one, its location preserves.
One should take into account that, in accordance with our definition of phase shift $\delta$, the LPT surrounding the zone-boundary stationary point $(\theta=\pi/4, \delta=0)$ corresponds to equal velocities of the modes and the trajectory encircling another stationary point  $(\theta=\pi/4,\delta=\pi)$ - to equal displacements. Therefore, the initial conditions in the MD experiments correspond to  fixed velocity and zeroth initial displacements of carbon atoms. This circumstance has to be taken into account while interpreting both experimental and MD simulation data.

We demonstrate that instability of edge-spectrum optical modes of CNT vibrations is the preliminary condition of energy localization in the some domain of CNT. The energy capture in one of the CNT parts can be achieved, if the excitation level exceeds the specific threshold, which corresponds to merging two trajectories - the LPT and the separatrix. When this  threshold is exceeded  the phase portrait of the system under consideration changes drastically: the separatrix passing through the unstable stationary point ($\theta=\pi/4$, $\delta=0$) (see fig. \ref{fig:largeX}(b)) encircles the stable stationary point ($\theta=\pi/4$, $\delta=\pi$) and prevents full energy exchange between effective particles $\psi_{1}$ and $\psi_{2}$. Simultaneously a set of transit-time trajectories, which involve any values of phase difference $\delta$, is created. It means that initial conditions corresponding to identical velocities or displacements  of both  modes lead to the energy capture by the effective particles. Then only partial energy exchange becomes possible along the trajectories, surrounding the stable stationary point and situated inside the separatrix. One should keep in mind that this process of energy capture does not suggest the creation of strongly localized solutions whose formation requires a participation of more  components of the spectrum. This can be achieved for CNT with larger aspect ratio.

As a conclusion we would like to note that the phenomenon of energy localization considered above has universal character and it is the common peculiarity of the systems possessing the optical-type branches of vibrational spectrum. In particular, one can assume that such well-known optical-type mode as RBM can manifest the energy capture in the effective particles too. The main difference is in the type of localization. Because we assume the hard nonlinearity in the RBM branch one should expect a "dark"-type localized solution in contrast with CFM branch, where nonlinearity is of soft type. We hope that these processes can be observed experimentally.

\begin{acknowledgments}
The work was supported by Program of Department of Chemistry and Material Science (Program \#1), Russia Academy of Sciences, and Russia Basic Research Foundation (grant 08-03-00420a)
\end{acknowledgments}

\bibliography{SmirnovVV_2013x}
%
%

\end{document}